\documentclass[12pt, draftclsnofoot, onecolumn]{IEEEtran}
\usepackage{amsmath}
\usepackage{amssymb}
\usepackage{amsthm}
\usepackage{cite}
\usepackage{color}
\usepackage{xcolor}
\usepackage{citesort}
\usepackage{caption}
\usepackage{epsfig,latexsym}
\usepackage{float}
\usepackage{fancyhdr}
\usepackage{graphicx}
\usepackage{indentfirst}
\usepackage{mdwmath}
\usepackage{mdwtab}
\usepackage{subfigure}
\usepackage{setspace}
\usepackage{times}
\usepackage{url}
\usepackage{optidef}
\usepackage{algorithm}
\usepackage{algorithmic}
\usepackage{listings}
\usepackage{bm}

\newtheorem{remark}{Remark}
\newtheorem{theorem}{Theorem}

\newtheorem{lemma}{Lemma}

\newtheorem{corollary}{Corollary}

\newtheorem{proposition}{Proposition}

\begin{document}
\title{Intelligent Reflecting Surface Aided Multiple Access Over Fading Channels}
\author{Yiyu Guo,
        Zhijin Qin,~\IEEEmembership{Member,~IEEE,} 
        Yuanwei Liu,~\IEEEmembership{Senior Member,~IEEE,} 
        and Naofal Al-Dhahir,~\IEEEmembership{Fellow,~IEEE,}

\thanks{Y. Guo, Z. Qin and Y. Liu are with School of Electronic Engineering and Computer Science, Queen Mary University of London, London E1 4NS, U.K. (e-mail: yiyu.guo@qmul.ac.uk; z.qin@qmul.ac.uk; yuanwei.liu@qmul.ac.uk).}
\thanks{N. Al-Dhahir is with the Department of Electrical and Computer Engineering, The University of Texas at Dallas, Richardson, TX 75080 USA (e-mail: aldhahir@utdallas.edu).}
}

\maketitle

\begin{abstract}
This paper considers a two-user downlink transmission in intelligent reflecting surface (IRS) aided network over fading channels. Particularly, non-orthogonal multiple access (NOMA) and two orthogonal multiple access (OMA) schemes, namely, time division multiple access (TDMA) and frequency division multiple access (FDMA), are studied. The objective is to maximize the system average sum rate for the delay-tolerant transmission. We propose two adjustment schemes, namely, dynamic phase adjustment and one-time phase adjustment. The power budget, minimum average data rate, and discrete unit modulus reflection coefficient are considered as constrains. To solve the problem, two phase shifters adjustment algorithms with low complexity are proposed to obtain near optimal solutions. With given phase shifters and satisfaction of time-sharing condition, the optimal resource allocations are obtained using the Lagrangian dual decomposition. The numerical results reveal that: i) the average sum rate of proposed NOMA network aided by IRS outperforms the conventional NOMA network over fading channels; ii) with continuous IRS adjustment in the fading block, the proposed TDMA scheme performs better than the FDMA scheme; iii) increasing the minimum average user rate requirement has less impact on the proposed IRS-NOMA system than on the IRS-OMA system. 
\end{abstract}


\section{Introduction}

\IEEEPARstart{H}{igher} demands on spectrum efficiency, energy efficiency, and massive connectivity have been placed for the fifth generation (5G) and beyond wireless communication systems\cite{dai2015non,lien20175g,letaief2019roadmap}. Intelligent reflecting surface (IRS) technology has gained extensive attention recently. Specifically, IRS is a metasurface consisting of multiple passive reflecting elements, which could dynamically change the signal direction\cite{mu2019exploiting,liang2019large}. Its ability of controling the signal reflections and the propagation leads to the performance gain with low power and cost\cite{gong2019towards}. As a promising technology for smart radio environments, IRS brings benefits to scenarios like smart cities, homes, and stations, with the potential applications like coverage enhancement, interference suppression and energy efficiency. The metasurfaces can be configured in order to create configurable links in dead-zones and to steer signals towards specific directions or locations for enhancing the signal quality and suppressing unwanted signals that may interfere, with no power amplification after configuration. Moreover, the transparent substrate does not interfere aesthetically or physically with the surrounding environment or with the line-of-sight of people, thus making the structure suitable on buildings, vehicles, and billboards\cite{di2020smart}.

IRS-assisted communication shares the similarity with the amplifying and forwarding (AF) relay communication and backscatter communication, but also with some important differences\cite{di2020smart,el2019delay,jung2020performance}. The AF relay requires active signal processing and retransmits the amplified signal, while the IRS elements only reflect signals passively without introducing any active processing of them\cite{tang2019wire,li2020reconfigurable}. Hence, IRS can greatly reduce energy consumption\cite{ozdogan2019intelligent}. This feature also enables the IRS to work in full-duplex mode, which can avoid the self-interference in comparison with the full-duplex AF relay\cite{huang2019reconfigurable}. For the backscatter communication, either a dedicated source or ambient source is needed for backscatter communication\cite{yang2017modulation,zhang2018constellation,long2019full,guo2020weighted}. Backscatter devices transmit its own information and modulate the incident signal, while the IRS provides an additional link without introducing its own information\cite{jung2019reliability}. 

To fully achieve the passive beamforming gains of IRS, multiple access techniques are of significant importance\cite{zhang2019capacity}. Based on design principles, multiple access schemes can be classified as orthogonal multiple access (OMA) and non-orthogonal multiple access (NOMA) \cite{al2019energy}. Conventional OMA communication orthogonally serves one user in a single resource block in the time/frequency/code domains or the combinations thereof\cite{hou2018multiple}. By contrast, the successive interference cancellation (SIC) technique enables the power domain NOMA scheme to achieve a higher number of connections in one resource block \cite{liu2017non}. The features of NOMA scheme have been widely studied recently\cite{liu2017nonorthogonal,shirvanimoghaddam2017massive,ding2015impact}.

\subsection{Related Works}

The study and design of IRS-assisted networks have focused towards new challenges of IRS passive beamforming\cite{perovic2019channel,ntontin2019multi}. Various wireless communication networks assisted by the IRS have been investigated for verifying its performance gains\cite{basar2019transmission}. An IRS aided multiuser multiple-input single-output (MISO) system was investigated for the single cell case in \cite{wu2019intelligent}. By optimizing IRS elements reflection coefficients, the power consumption can be significantly reduced. The IRS was utilized for secure transmission relying on cooperative jamming techniques in \cite{han2019intelligent,guan2019intelligent}. The system fairness was considered in \cite{nadeem2020asymptotic}, neglecting the direct link for blockage, and the users' minimum rate was maximized with IRS assisted. The IRS has also been used to enhance the rate performance in \cite{yang2019intelligent}. 

Beside the works aforementioned under the OMA schemes, recently, some researchers have investigated the efficient integration of IRS with NOMA to enhance spectrum and energy efficiencies\cite{liu2017enhancing,mu2020capacity}. A SISO-NOMA IRS-assisted network was proposed in \cite{hou2019reconfigurable}, where a prioritized design was proposed for further enhancing spectrum and energy efficiencies. To solve the beamforming and IRS design problems, a difference-of-convex (DC) algorithm was applied in \cite{fu2019reconfigurable}. By considering the target data rate constrains, decoding order problem was also investigated. The decoding order problem was also studied in \cite{yang2019intelligent} and users were ordered base on the combined channel gains. The authors in \cite{yue2020performance} proposed a 1-bit coding scheme on the IRS, which transmits signals to the NOMA users as a relay. With the ability of adjusting the signal directions, the IRS has been regarded as a promising technology that illustrates important advances with the implement in NOMA networks\cite{ding2019simple}. With the goal of minimizing the transmit power under the discrete unit-modulus constraint on the magnitude of each IRS element, Zheng et al.\cite{zheng2020in} analyed the comparison of NOMA and OMA in IRS-aided downlink communication with two users. Their theory showed that the TDMA scheme may perform better than the NOMA scheme when the NOMA users have symmetric rates and deployment.

Moreover, dynamic resource allocation over fading channels significantly enhances the performance compared with fixed resource allocation in static channels\cite{xing2018optimal}. In systems with fading broadcast channels, power, time, and bandwidth are adaptively assigned among users based on the channel state information (CSI). Particularly, the dynamic resource assignment policies under various multiple access schemes were proposed in \cite{li21capacity} and \cite{li22capacity}, which assumed that the perfect CSI is available. The corresponding ergodic capacity region and the outage capacity region were studied as well. By utilizing spectrum sensing and sharing, a power assignment policy was investigated in \cite{asghari2011resource} for maximizing the achievable capacity over fading channels.

\subsection{Motivations and Contributions}

As aforementioned, the IRS has been researched in various scenarios, but is still in its infancy. Although IRS requires no power for amplifying or processing the signal, the power dissipation and hardware loss is non-negligible over the long term for the elements adjustment. In particular, the distinct difference of users received power levels are critical for determining the decoding order and performance in the NOMA scheme, which can be influenced by IRS over fading channels.

Most works on IRS-assisted systems were under static channels, which motivate us to develop a long-term resource allocation policy for IRS-assisted downlink networks over fading channels. The transmit power, time/frequency resources, and phase shift should be adaptively allocated under fading broadcast channels in all fading states, to maximize the average sum rate.

In this paper, both OMA and NOMA schemes are investigated for comparison. Two types of OMA schemes, namely, frequency division multiple access (FDMA) and time division multiple access (TDMA), are considered. To our best knowledge, such a dynamic optimization problem with multiple access schemes over fading channels scenario is still lack of studies. The main challenges of the considered system are identified as the following: 
\begin{itemize}
\item The formulated resource allocation problem is non-trivial to solve as the objective functions is NP-hard.
\item The global optimal solutions is hard to obtain by efficient methods or algorithms, as the discrete unit-modulus constraints and the non-convexity lies in the optimization problems.
\item The long term design is expected for the coupled IRS adjustment and resource allocation problems over fading channels.
\end{itemize}

In sight of the above challenges, we consider the joint phase shifters policy and resource assignment optimization problems in downlink IRS-assisted systems. The major contributions are outlined below: 
\begin{enumerate}
\item A downlink IRS-aided system over fading channels is considered, where two users are served by different multiple access techniques, namely, NOMA, TDMA and FDMA. We propose two IRS adjustment schemes based on the timing of altering reflection coefficients. The joint optimization of phase shift and resource allocation is solved for the average sum rate maximization.
\item For the IRS design, we apply the success convex approximation (SCA) to solve the IRS adjustment problems, where the sequential rank-one constraint relaxation (SROCR) method is applied for the rank-one constrains. Low complexity algorithms are proposed for IRS phase adjustment. 
\item For the resource allocation, we utilize the Lagrange duality method for the non-convex problems. The formulated average sum rate maximization problems for all multiple access schemes meet the time-sharing condition, then adaptive solutions that achieve maximum data rate over fading channels are obtained.
\item Alternating optimization (AO) algorithms are proposed to optimize IRS reflection coefficients and power allocation efficiently. The proposed communication networks can enhance the system effectiveness by integrating IRS and NOMA. Numerical results demonstrate the utility of the proposed methods.

\end{enumerate}

\subsection{Organization and Notations}

The remainder of this paper is organized as follows. In Section II, the system model is presented. In Section III, two IRS adjustment policies and phase shifters optimization problems are formulated first. Then the proposed design to solve the resource allocation problem for the IRS-aided network is presented in detial. We then provide the simulation results to verify the theoretical analyses in Section IV. The conclusions follows in Section V. 

The notations used are shown as follows. $\mathbb{C}^{N \times N}$ denotes the space of $N \times N$ complex-valued matrices and $diag(\cdot)$ denotes diagonal elements of a matrix. $x^H$ denotes the complex conjugate transpose of vector $x$. The notations $Tr(\cdot)$ and $rank(\cdot)$ denote the trace and rank of a matrix, respectively. $\mathcal{R}e(x)$ and $\mathcal{I}m(x)$ denote the real and imaginary parts of the complex number x, respectively. $\left [ x \right ]^b_a $ denotes the value of $ max(min(x,b),a)$, and $arg(\cdot)$ denotes the component-wise phase of a complex vector.

\section{System Model}

As shown in Fig.1, the considered downlink system consists of one base station (BS) with the single antenna, IRS with $N$ passive reflecting elements and $K$ users equipped with the single antenna. All channels are assumed that consist of large scale and small scale fading. The channel between BS and user k is denoted as $h_k$. The small scale fading between BS and user k is modeled as Rayleigh fading, then the corresponding channel can be expressed as 
\begin{equation}\label{rayleigh}
h_k =\sqrt{L(d)} f^{\text{NLos}},
\end{equation}
where $f^{\text{NLos}}$ is the Rayleigh fading. The large scale pass loss is modeled as $L(d) =\rho_0(\frac{d}{d_0})^{-\varphi}$, where $\rho_0=-30dB$, and $\varphi$ denotes the path loss exponent. 

For the channel between BS and IRS and that between IRS and users, line of sight (LoS) components exist. Therefore, the small scale fading effects in these channels are modeled as Rician fading. The channels from BS to IRS and IRS to users are denoted as $g \in \mathbb{C}^{N \times 1}$ and $r_k \in \mathbb{C}^{N \times 1} $, respectively. The resulting BS to IRS channel is following:
\begin{equation}\label{rician}
g = \sqrt{L(d)} (\sqrt{\frac{v}{1+v}}f^{\text{Los}} + \sqrt{\frac{1}{1+v}}f^{\text{NLos}}),
\end{equation}
where $v$ is Rician factor and $f^{\text{Los}}$ refers to the LoS component. The IRS to users channels are modeled in the similar way. 

We assume that the complex channel coefficients $h_k(i)$, $g(i)$, and $r_k(i)$ experience block fading with a continuous joint probability density function (pdf), where $i$ represents a fading state. In this paper, based on the assumption that the IRS elements can only be modified by adjusting the phase, the IRS behaves as an anomalous mirror on the order of tens of meters \cite{di2020r,di2020analytical}. Define $\bm{\Theta} = diag(u) \in \mathbb{C}^{N \times N}$ as the IRS diagonal reflection coefficients matrix with $u = [ u_1, u_2, ..., u_N]$ and $u_n = \beta_ne^{j\theta_n}$, where $\beta_n \in [0,1]$ and $\theta_n \in [0,2\pi)$ refer to the amplitude of the reflection coefficient and phase shift of the $n$th IRS element, respectively. The IRS phase quantization level is denoted as $L$ and $\Delta{\theta}=2\pi/{2^L}$. Then, the phase shift is discrete with $\mathcal{D} = \{0, \Delta{\theta}, ..., (2^L-1)\Delta{\theta} \}$.

Two multiple access schemes are considered, namely, NOMA and OMA. For the NOMA scheme, the two users share the same resource block simultaneously. For the OMA scheme, continuous time/frequency allocations are considered.

1) NOMA: A power-domain downlink NOMA scheme is implemented in the considered system. The signal transmitted to user 1 is given by
\begin{equation}\label{receive signal,eq1}
y_1(i)=(h_1(i)+r_1^H(i)\bm{\bm{\Theta}} g(i))(\sqrt{p_1(i)}s_1+\sqrt{p_{2}(i)}s_{2})+n_1,
\end{equation}
where $s_1$ is the transmit signal intended for user 1 and $\mathbb{E}\{ s_1^2 \}=1 $; $p_1(i)$ denotes the transmit power for user 1; $n_1 \sim \mathcal{CN}(0, \sigma^2)$ is the additive white Gaussian noise (AWGN) with zero mean and variance $\sigma^2$.

According to the NOMA decoding principle, the user with poor channel gain is assigned with a higher power budget and that with better channel condition performs SIC to remove the signal of the other user. Proper decoding order based on the channel conditions and transmit power levels can significantly enhance the NOMA transmission performance. In IRS-aided NOMA networks, the combined channel gain is determined by both the diret link and the auxiliary IRS link, $h_k(i), r_k^H(i)$ and $g(i)$. As $\bm{\Theta}$ can be manually controled, IRS significantly complicates user decoding order in NOMA scheme. With two users served in the system, we will consider all possible SIC decoding orders, denoted as $\Pi$. \\
\indent Then, the signal-to-interference-plus-noise ratio (SINR) for user 1 at fading state $i$ can be given by
\begin{equation}\label{rate,eq3}
\tau_1^{\text{NOMA}}(i) =
\begin{cases}
\frac{\vert h_1(i)+r_1^H(i)\bm{\Theta} g(i) \vert^2 p_1(i)}{\vert h_1(i)+r_1^H(i)\bm{\Theta} g(i) \vert^2 p_{2}(i) + \sigma^2}, &\text{if } \Omega (1)(i) =1,\\
\frac{\vert h_1(i)+r_1^H(i)\bm{\Theta} g(i) \vert^2 p_1(i)}{\sigma^2}, &\text{otherwise,}
\end{cases}
\end{equation}
where $\Omega (k)$ denotes the user k's ordering and $\Omega (k)= 1$ means that user k's signal will be decoded first. Then, the instantaneous achievable rate for user 1 is given by $R_{1}(i)=log_2(1+\tau_{1}^{\text{NOMA}}(i))$.

\begin{figure}[t!]
\centering
\includegraphics[width= 2.7in, height=2.2in]{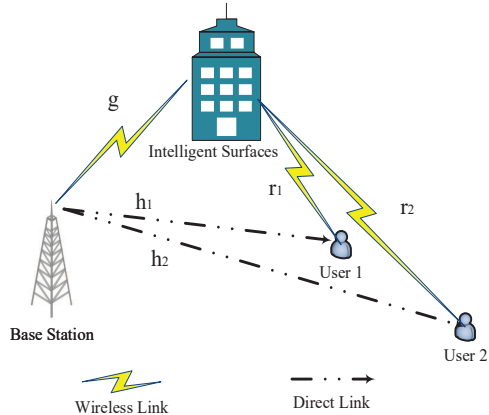}
\caption{Illustration of an IRS-assited downlink NOMA network.}
\label{system_model}
\end{figure}

2) OMA:  For the OMA scheme, the BS serves users in orthogonal frequency bands and time slots under FDMA and TDMA schemes, respectively. At the fading state $i$, the instantaneous achievable rate for user k is given by
\begin{equation}\label{sinro,eq4}
R_k^{\text{OMA}}(i)= \alpha_k(i)log_2(1+ \frac{\vert h_k(i)+r_k^H(i)\bm{\Theta} g(i) \vert^2 p_k(i)}{\alpha_k(i)\sigma^2} ),
\end{equation}
where $\alpha_k(i)$ denotes a fraction of the orthogonal resource at the fading state $i$, and $\alpha_1(i)+\alpha_2(i)=1$ with $\alpha_k(i) \in [0,1]$. The total power consumption in the FDMA scheme is same as that in the TDMA scheme, i.e., $\alpha_1(i) \frac{p_1(i)}{\alpha_1(i)}  + \alpha_2(i) \frac{p_2(i)}{\alpha_2(i)} = p_1(i)+p_2(i), \forall i$.

\section{Joint Phase Shift Design and Power Allocation}

\subsection{Problem Formulation}
We consider the transmissions in delay-tolerant networks, which is tolerant with the prorogation delay. Thus the number of codewords can be designed for a length that goes to infinity in theory. The users decode their desired signals from the BS in the entire fading procedure. Then, the average sum rates in fading state $i$ are given by $\mathbb{E}_i[R_1^{\text{NOMA}}(i)+R_{2}^{\text{NOMA}}(i)]$ and $\mathbb{E}_i[R_1^{\text{OMA}}(i)+R_{2}^{\text{OMA}}(i)]$ for the NOMA and OMA schemes, respectively. 

Each fading state lasts a very short time, which makes it costly to adjust the intelligent surfaces continuously. Therefore, we consider two schemes for adjusting the phase shift.
\begin{itemize}
\item \textbf{Dynamic phase adjustment:} In this scenario, the IRS adjustment can be performed in every fading state based on the instantaneous transmit power for each user and channel state information obtained from the previous fading state. The transmit power, time and frequency allocation will be performed under given phase shifts.

\item \textbf{One-time phase adjustment:} In this scenario, we divide all fading states into several blocks. The reflecting elements are adjusted at the beginning of each fading block based on the average transmit power for each user and CSI obtained from the previous block. The IRS in TDMA scheme will be optimised in the time slot same as that in the FDMA scheme for this scenario.
\end{itemize}

\begin{remark}\label{re1}
There is a major difference between the FDMA and TDMA schemes under dynamic phase adjustment, namely that the IRS can be adjusted for a single user in TDMA transmission at different time slots while it needs to be optimised for both users to maximise the ergodic sum-rate in FDMA transmission. 
\end{remark}

In delay-tolerant networks, the user with better channel condition is allocated with most resource budget for maximing the average sum-rate over the entire fading process, which motivates the designer to set a target rate for the poor user. Aiming at data rate maximization while guaranteeing the user fairness, we jointly optimize the phase shift and resources allcation at each fading state or fading block. The optimization problem with the NOMA scheme can be formulated as
\begin{maxi!}|s|
{p_1(i),p_2(i),\bm{\Theta}}{\mathbb{E}_i[R_1^{\text{NOMA}}(i)+R_{2}^{\text{NOMA}}(i)]}
{}{\textbf{(P1)}}
\addConstraint{\mathbb{E}_i[p_1(i) +p_{2}(i)] \leq \overline{P}  \label{objective:c1} } 
\addConstraint{p_1(i) +p_{2}(i) \leq \hat{P}, \forall i          \label{objective:c2} }
\addConstraint{p_1(i) \geq 0, p_2(i) \geq 0, \forall i           \label{objective:c3} }
\addConstraint{\mathbb{E}_i[R_k(i)] \geq \overline{R}, \forall k \label{objective:c4} }
\addConstraint{\theta_{n} \in \mathcal{D},\quad \forall n = 1, ...,N  \label{objective:c5} }
\addConstraint{\Omega \in \Pi, \label{objective:c6} }{}
\end{maxi!}
where $\overline{P}$ denotes the average power budget that is determined by the total transmit power over the long term. $\hat{P}$ denotes the peak power constraint determined by the transmit power budget at the fading state $i$. Constraint \eqref{objective:c1} describes the average transmit power budget during the entire transmission procedure. Constraint \eqref{objective:c2} limits the instantaneous total transmit power at one fading state. We can see that $\overline{P} \leq \hat{P}$. Constraint \eqref{objective:c4} limits the minimum transmission rate for each user. Constraint \eqref{objective:c5} limits the IRS reflection coefficients in the certain discrete values in $\mathcal{D}$, which will be neglected in the first place and obtained by quantizing to its nearest discrete value in $\mathcal{D}$ from the continuous solutions. Constraint \eqref{objective:c6} represents the combination set of all possible decoding orders.

The direct links could be weak for the blockage or distance, which is adverse to the performance of NOMA scheme. While the IRS can help to establish the combine channel which is strong and makes the implements of NOMA scheme possible. The links in IRS-aided system are configurable and it should be noted that the phase response will affect the gain of the combined channel and the difference between the users' received power levels, which is critical to determining the decoding order and performance in the NOMA scheme. In this paper, the direct link exists as a part of the combined channels. Hence, the blocked case can be easily obtained by neglecting the direct link.

For the OMA scheme, the optimization problem is formulated as
\begin{maxi!}|s|
{p_1(i),p_2(i),\bm{\Theta}}{\mathbb{E}_i[R_1^{\text{OMA}}(i)+R_{2}^{\text{OMA}}(i)]}
{}{\textbf{(P2)}}
\addConstraint{\eqref{objective:c1}, \eqref{objective:c2}, \eqref{objective:c3}, \eqref{objective:c4}, \eqref{objective:c5}.}{}
\end{maxi!}

\subsection{Phase Shift Adjustment}

\subsubsection{\text{NOMA}} With given transmit power in corresponding phase adjustment schemes, we can obtain the phase shifts $\bm{\Theta}$ for the NOMA scheme by solving
\begin{maxi!}|s|
{\bm{\Theta}}{\mathbb{E}_i[R_1^{\text{NOMA}}(i)+R_{2}^{\text{NOMA}}(i)]}
{}{\textbf{(P1.1)}}
\addConstraint{\eqref{objective:c4}, \eqref{objective:c5}.}{}
\end{maxi!}
The problem is a complicated optimization problem. The non-convexity is presented by its non-convex objective function and a discrete unit-modulus constraint on reflecting elements. The globally optimal solution is hard to obtain by efficient methods or tools. One way to reach the globally optimal phase shifts is exhaustive search. Specifically, the sum rates under all possible cases of $\Theta$ are calculated and the one that achieves the maximum sum rate is chosen as the optimal solution. However, brute-force search method can lead to exponential complexity with $O(L^N)$, which is time consuming if N and L are large. Therefore, we only use the exhaustive search as the upper bound for our proposed methods. \\
\indent Considering aforementioned reasons, we propose efficient algorithms that can achieve the near optimal performence with low computational complexity. In particular, the alternating optimization (AO) method is utilized. We first drop the discrete constraints and replace them with continuous phase shifts. We quantize each phase shift after obtaining the continuous-valued solutions. To this end, let $I_k=diag(r_k^H)g$, $\overline{u}=[u, 1]$, and we introduce
\begin{equation}\label{rate,eq00}
Z_k = \begin{bmatrix} I_k \\ h_k^H \end{bmatrix}.
\end{equation}
Then, we can obtain $\vert h_k+r_k^H\bm{\Theta} g \vert^2 = \vert \overline{u}Z_k \vert^2 $. Let $\frac{1}{X_{kj}}=\vert \overline{u}Z_k \vert^2 p_j$ and $ Y_{kj} =\vert \overline{u}Z_k \vert^2 p_j + \sigma^2$. Given the decoding order, i.e., $\Omega (1) < \Omega (2)$, problem (P1.1) can be transformed into a more tractable form
\begin{maxi!}|s|
{\bm{\Theta}}{\mathbb{E}_i[R_1^{\text{NOMA}}(i)+R_{2}^{\text{NOMA}}(i)]}
{}{\textbf{(P1.1.1)}}
\addConstraint{log_2(1+\frac{1}{X_{11}Y_{12}}) \geq \overline{R} \label{objective:e1} }
\addConstraint{log_2(1+\frac{1}{X_{22}\sigma^2}) \geq \overline{R} \label{objective:e2} }
\addConstraint{\frac{1}{X_{kk}} \leq \vert \overline{u}Z_k \vert^2 p_k, \quad k=1,2 \label{objective:e3} }
\addConstraint{Y_{12} \geq \vert \overline{u}Z_1 \vert^2 p_2 + \sigma^2 \label{objective:e4} }
\addConstraint{\vert \overline{u}_n \vert^2 = 1, \forall n,  \label{objective:e5} }{}
\end{maxi!}
where \eqref{objective:e3} and \eqref{objective:e4} serve as lower bounds on $X_{kj}$ and $ Y_{kj}$. The constraints \eqref{objective:e1}, \eqref{objective:e3}, and \eqref{objective:e5} are non-convex. Note that $log_2(1+\frac{1}{XY})$ is a joint convex function with respect to $X$ and $Y$. Then, the lower bound at given local points $\{X(l), Y(l) \}$ can be achieved by utilizing the first-order Taylor expansion. Correspondingly, the lower bound of the weak user and strong user can be expressed as 
\begin{equation}\label{rate,sca1}
\begin{split}
&log_2(1+\frac{1}{X_{11}Y_{12}}) \geq R_{1}^{low},
\end{split}
\end{equation}
and 
\begin{equation}\label{rate,sca2}
\begin{split}
&log_2(1+\frac{1}{X_{22}\sigma^2}) \geq R_{2}^{low},
\end{split}
\end{equation}
respectively, where
\begin{equation}\label{r1l}
\begin{split}
R_{1}^{low}& = log_2(1+\frac{1}{X_{11}(l)Y_{12}(l)})- \\ &\frac{(log_2^e)(X_{11}-X_{11}(l))}{X_{11}(l)+X_{11}(l)^2Y_{12}(l)}-\frac{(log_2^e)(Y_{12}-Y_{12}(l))}{Y_{12}(l)+Y_{12}(l)^2X_{11}(l)},
\end{split}
\end{equation}
and
\begin{equation}\label{r2l}
\begin{split}
R_{2}^{low}& =log_2(1+\frac{1}{X_{22}(l)\sigma^2})-\frac{(log_2^e)(X_{22}-X_{22}(l))}{\sigma^2(X_{22}(l)+X_{22}(l)^2\sigma^2)}.
\end{split}
\end{equation}
Then the term on the right side of \eqref{objective:e3} is a convex function with respect to $\overline{u}$. Similarly, at the given local point $\overline{u}(l)$, the lower bound achieved by utilizing first-order Taylor expansion can be given by
\begin{equation}\label{rate,sca12}
\vert \overline{u}Z_k \vert^2 \geq \delta_{k}= \vert \overline{u}(l)Z_k \vert^2 + 2\mathcal{R}e( (\overline{u}(l)Z_kZ_k^H)(\overline{u}-\overline{u}(l))).
\end{equation}
Thus, the phase shifters adjustment problem can be approximated as
\begin{maxi!}|s|
{\bm{\Theta}}{\mathbb{E}_i[R_1^{\text{NOMA}}(i)+R_{2}^{\text{NOMA}}(i)]}
{}{\textbf{(P1.1.2)}}
\addConstraint{R_{k}^{low} \geq \overline{R}, \quad k=1,2 \label{objective:f1} }
\addConstraint{\frac{1}{X_{kk}} \leq \delta_{k}p_k, \quad k=1,2 \label{objective:f3} }
\addConstraint{Y_{12} \geq \vert \overline{u}Z_1 \vert^2 p_2 + \sigma^2 \label{objective:f4} }
\addConstraint{\vert \overline{u}_n \vert^2 = 1, \forall n, \overline{i}_{n+1}=1.  \label{objective:f5} }{}
\end{maxi!}
\indent One common method for rank-one constraint \eqref{objective:f5} is applying semidefinite relaxation (SDR), where the rank-one constraint is first ignored and a solution with random rank is obtained. Then the solution is constructed to the rank-one form by applying Gaussian randomization method. Although dropping the rank-one constraint could lead to a relaxed convex problem, the obtained solution is usually sub-optimal. To overcome this challenge, a novel algorithm based on sequential rank-one constraint relaxation (SROCR) is applied.
\begin{maxi!}|s|
{\bm{\Theta}}{\mathbb{E}_i[R_1^{\text{NOMA}}(i)+R_{2}^{\text{NOMA}}(i)]}
{}{\textbf{(P1.1.3)}}
\addConstraint{R_{k}^{low} \geq \overline{R}, \quad k=1,2 \label{objective:z1} }
\addConstraint{\frac{1}{X_{kk}} \leq Tr(UZ_kZ_k^H)p_k, \quad k=1,2 \label{objective:z3} }
\addConstraint{Y_{12} \geq Tr(UZ_1Z_1^H)p_2 + \sigma^2 \label{objective:z4} }
\addConstraint{U \succeq 0  \label{objective:z5} }
\addConstraint{rank(U)=1    \label{objective:z6} }
\addConstraint{[U]_{nn}=1.   \label{objective:z7} }{}
\end{maxi!}
where $U=\overline{u}\overline{u}^H$, $ U \succeq 0$, $rank(U)=1$ and $[U]_{nn}=1$. We first apply a partial constraint relaxation for the constraint \eqref{objective:z6} and the relaxed problem (P1.1.4) is given by
\begin{maxi!}|s|
{\bm{\Theta}}{\mathbb{E}_i[R_1^{\text{NOMA}}(i)+R_{2}^{\text{NOMA}}(i)]}
{}{\textbf{(P1.1.4)}}
\addConstraint{e_{max}(U^{(i)})^HUe_{max}(U^{(i)}) \geq \kappa^{(i)}Tr(U)  \label{objective:zx1} }
\addConstraint{\eqref{objective:z1}, \eqref{objective:z3}, \eqref{objective:z4}, \eqref{objective:z5},  \label{objective:zx2} }{}
\end{maxi!}
where $\kappa^{(i)} \in [0,1]$ is the relaxation parameter, $U^{(i)}$ denotes the largest eigenvalue of $U$ and $e_{max}(U^{(i)})$ denotes the eigenvector of the $U^{(i)}$. $\kappa^{(i)}$ controls the largest eigenvalue to trace ratio of $U$, as the solution $U^{(i)}$ in the i-th iteration. The problem $\text{(P1.1.4)}$ is a convex problem, which can be solved efficiently by the standard convex optimization tools, such as CVX. When $\kappa^{(i)} = 0$, constraint \eqref{objective:zx1} could be considered as dropping the rank-one constraint, while $\kappa^{(i)} = 1$, constraint \eqref{objective:zx1} approaches to the original rank-one constraint in \eqref{objective:z6}. Note that due to the relaxation replacement, the solution of problem $\text{(P1.1.4)}$ serves as the lower bound of problem $\text{(P1.1)}$. Then, we apply Cholesky decomposition, e.g. $U^{l+1}=\overline{u}\overline{u}^H$, to find the phase shifters. Algorithm 1 summarizes the proposed algorithm to solve problem (P1.1.3).

\begin{algorithm}
\caption{Proposed algorithm for NOMA phase adjustment}
\begin{algorithmic}[1]
\STATE \textbf{Initialize:} Convergence thresholds $\epsilon_1$, $\epsilon_2$, a feasible solution $U(l)$, step size $\Delta^{(l)}$ and iteration index $l=0$.
\STATE Solve problem (P1.1.4) and obtain $u^{(l)}$ with $\kappa(l) = 0$. 
\REPEAT 
\STATE Solve problem (P1.1.4) with $\{ \kappa^{(l), U^{(l)}} \}$.
\STATE if $\{ \kappa^{(l), U^{(l)}} \}$ is feasible \textbf{then}
\STATE Obtain the optimal solution $\overline{U}^{(l+1)}$, 
\STATE $\Delta^{(l+1)}=\Delta^{(l)}$;
\STATE \textbf{else}
\STATE $\Delta^{(l+1)}=\Delta^{(l)}/2$;
\STATE \textbf{end}
\STATE $\kappa^{(l+1)} = min (1 , \frac{e_{max}(U^{(l+1)})}{Tr(U^{(l+1)})}+\Delta^{(l+1)}).$
\STATE $l=l+1$.
\UNTIL $\kappa^{(l-1)} \geq \epsilon_1$ and objective value with the obtained $U$ reaches convergence with $\epsilon_2$.
\end{algorithmic}
\end{algorithm}

\begin{theorem}
\emph{With initial solution of phase shifts, Algorithm 1 converges to a KKT stationary point of problem (P1.1.4), which is equivalent to problem (P1.1.3).}
\begin{proof}
The algorithm to build a rank one solution can be considered as an AO of $u = e_{max}(U^{(i)})$ and $U$. The proof is detailed in \cite{cao2017sequential}.
\end{proof}
\end{theorem}

The obtained solution $u$ is still continuous. The nearest feasible discrete phase shifters are obtained by quantizing as
\begin{equation}\label{rate,scaf}
u^* =
\begin{cases}
e^{j\theta_n^*}, &  n=1,...,N,\\
1 & n=N+1,
\end{cases}
\end{equation}
where 
\begin{equation}\label{rate,scaf1}
\theta_n^*=arg \min \limits_{\theta \in \mathcal{D}} \vert \theta-\theta_n  \vert.
\end{equation} 

Note that the newly obtained solution $u^*$ may not be optimal, thus solution $u^*$ is updated only when the objective function’s value increases. In the following, two types of multiple access schemes, namely. FDMA and TDMA, are consider in both dynamic phase adjustment and one-time phase adjustment schemes. 

\subsubsection{\text{OMA}} 
Since the IRS is to be optimised for maximising the sum rate of two users at the same time under both the FDMA and TDMA schemes in the one-time phase adjustment scheme, the formulation becomes
\begin{maxi!}|s|
{\bm{\Theta}}{\mathbb{E}_i[R_1^{\text{OMA}}(i)+R_{2}^{\text{OMA}}(i)]}
{}{\textbf{(P2.1)}}
\addConstraint{\eqref{objective:c4}, \eqref{objective:c5}.}{}
\end{maxi!}

Similarly, we introduce the slack variables $\frac{1}{X_{kj}}=\vert \overline{u}W_k \vert^2 p_j$ and $U=\overline{u}\overline{u}^H$, then problem (P2.1) is reformulated as 
\begin{maxi!}|s|
{\bm{\Theta}}{\mathbb{E}_i[R_1^{\text{OMA}}(i)+R_{2}^{\text{OMA}}(i)]}
{}{\textbf{(P2.1.1)}}
\addConstraint{log_2(1+\frac{1}{X_{kk}\sigma^2}) \geq \overline{R}, \quad \forall k=1,2 \label{objective:of1} }
\addConstraint{\frac{1}{X_{kk}} \leq Tr(UZ_kZ_k^H)p_k, \quad k=1,2 \label{objective:of2} }
\addConstraint{U \succeq 0  \label{objective:of3} }
\addConstraint{rank(U)=1    \label{objective:of4} }
\addConstraint{[U]_{nn}=1.   \label{objective:of5} }{}
\end{maxi!}
By substituting \eqref{rate,sca1}, \eqref{rate,sca2} and \eqref{rate,sca12}, the passive beamforming optimization problem under the OMA scheme is approximated as
\begin{maxi!}|s|
{\bm{\Theta}}{\mathbb{E}_i[R_1^{\text{OMA}}(i)+R_{2}^{\text{OMA}}(i)]}
{}{\textbf{(P2.1.2)}}
\addConstraint{R_{k}^{low} \geq \overline{R}, \quad k=1,2 \label{objective:ofdm1} }
\addConstraint{\eqref{objective:of2}, \eqref{objective:of3}, \eqref{objective:of4}, \eqref{objective:of5}. \label{objective:ofdm3} }{}
\end{maxi!}

We also use the SROCR to solve the rank-one constraint as 
\begin{maxi!}|s|
{\bm{\Theta}}{\mathbb{E}_i[R_1^{\text{NOMA}}(i)+R_{2}^{\text{NOMA}}(i)]}
{}{\textbf{(P2.1.3)}}
\addConstraint{e_{max}(U^{(i)})^HUe_{max}(U^{(i)}) \geq \kappa^{(i)}Tr(U)  \label{objective:ox1} }
\addConstraint{\eqref{objective:of2}, \eqref{objective:of3}, \eqref{objective:of5}, \eqref{objective:ofdm1}.  \label{objective:ox2} }{}
\end{maxi!}

Now, problem (P2.1.3) is also convex, then the algorithm for phase adjustment under OMA scheme is summarized in Algorithm 2.
\begin{algorithm}
\caption{Proposed algorithm for OMA phase adjustment}
\begin{algorithmic}[1]
\STATE \textbf{Initialize:} Convergence thresholds $\epsilon_1$, $\epsilon_2$, a feasible solution $U(l)$, step size $\Delta^{(l)}$ and iteration index $l=0$;
\STATE Solve problem (P2.1.3) and obtain $u^{(l)}$ with $\kappa(l) = 0$.
\REPEAT 
\STATE Solve problem (P2.1.3) with $\{ \kappa^{(l), U^{(l)}} \}$.
\STATE if $\{ \kappa^{(l), U^{(l)}} \}$ is feasible \textbf{then}
\STATE Obtain the optimal solution $\overline{U}^{(l+1)}$, 
\STATE $\Delta^{(l+1)}=\Delta^{(l)}$;
\STATE \textbf{else}
\STATE $\Delta^{(l+1)}=\Delta^{(l)}/2$;
\STATE \textbf{end}
\STATE $\kappa^{(l+1)} = min (1 , \frac{e_{max}(U^{(l+1)})}{Tr(U^{(l+1)})}+\Delta^{(l+1)}).$
\STATE $l=l+1$.
\UNTIL $\kappa^{(l-1)} \geq \epsilon_1$ and objective value with the obtained $U$ reaches convergence with $\epsilon_2$.
\end{algorithmic}
\end{algorithm}

\subsubsection{\text{TDMA under dynamic phase adjustment scheme}} In the dynamic scheme, the IRS will be adjusted twice in one fading state to serve two users in the TDMA scheme, which motivates us to aim at maximising each user's concatenated channel response, $h_k(i)+r_k^H(i)\bm{\Theta} g(i)$, as the IRS could be optimised for one user in different time slots. IRS will be adjusted twice in one fading state to serve two users in the TDMA scheme.

Given the transmit power allocation, the problem (P1) satisfies the following inequality:
\begin{equation}\label{omath}
\vert h_k(i)+r_k^H(i)\bm{\Theta} g(i) \vert  \overset{(a)}\leq  \vert h_k(i)  \vert + \vert r_k^H(i)\bm{\Theta} g(i) \vert,
\end{equation}
where the equality in (a) holds if and only if $arg(r_k^H(i)\bm{\Theta} g(i))=arg(h_k(i))=\phi_0 $. We can always obtain a phase shifter $\bm{\Theta}$ that satisfies $(a)$ with equality. By changing the variables of $ r_k^H(i)\bm{\Theta} g(i) = v^Ha $, where $a=diag(r_k^H g(i))$, and dropping the constant term $h_k(i)$, we arrive at the following simplified problem:
\begin{maxi!}|s|
{v}{\vert v^Ha  \vert}
{}{\textbf{(P2.2)}}
\addConstraint{0 \leq \theta_n \leq 2\pi, \forall n \label{objective:o1} }
\addConstraint{arg(v^Ha)=\phi_0 .\label{objective:o2} }{}
\end{maxi!}

The optimal solution for problem (P2.2) can be obtained by
\begin{equation}\label{vtdms}
v^{*}=e^{j(\phi_0-arg(a))}=e^{j(\phi_0-arg(diag(r_k^H(i))g(i)))}. 
\end{equation}
The n-th phase shifter at the IRS can be expressed as
\begin{equation}\label{omap}
\begin{split}
\theta_n^{*} &= \phi_0 - arg(r_{n,k}^H(i)g_n(i)) \\
             &= \phi_0 - arg(r_{n,k}^H(i))- arg(g_n(i)),
\end{split}
\end{equation}
where $r_{n,k}^H(i)$ denotes the n-th element of $r_k^H(i)$, and $g_n(i)$ denotes the nth element of $g(i)$. Equation \eqref{omap} suggests the optimal phase shifts solution is that the reflection links and direct links is aligned with BS-user link to maximise the concatenated channel response for each user. Then, we optimize the power allocation as well as the time or frequency allocation in OMA schmems under given phase shifter $\bm{\Theta}$ in the following section.

\subsection{Power Allocation} 
\subsubsection{Optimal Solution to NOMA Scheme} With given phase shifter $\bm{\Theta}$ and minimum achievable rates $\overline{R}$, our objective is maximizing the system average sum rate in each fading state or fading block. We optimize the power allocation, subject to power budget constraints. The target rate for both users is also considered for ensure the fairness between the users. Thus the objective function is rewritten as follows 
\begin{maxi!}|s|
{p_1(i),p_2(i)}{\mathbb{E}_i[R_1^{\text{NOMA}}(i)+R_{2}^{\text{NOMA}}(i)]}
{}{\textbf{(P1.2)}}
\addConstraint{\eqref{objective:c1}, \eqref{objective:c2}, \eqref{objective:c3}, \eqref{objective:c4}, \eqref{objective:c6}.}{}
\end{maxi!}

We fix the reflection coefficients and decoding order in each fading state or fading block, then we solve the power allocation problem. The objective function of problem (P1.2) is non-convex. Thus the optimal solution is hard to obtained by efficient methods. The problem (P1.2) satisfies the “time-sharing” condition, which is always satisfied when the number of fading states goes to infinity \cite{yu2006dual}. Strong duality holds when such condition is met. By using the Lagrangian duality, the duality gap between the original and the dual problems ecomes zero when “time-sharing” condition is satisfied by problem (P1.2). Hence, we can obtain the optimal solution of problem (P1.2) by solving its dual problem. Then, the Lagrange duality method can be utilized to obtain the solution of problem (P1.2). The Lagrangian function of the primal problem is given by
\begin{equation}\label{L1}
\begin{split}
\mathcal{L}_{dt}^{\text{NOMA}}( &\{ p_1(i) \}, \{ p_2(i) \}, \lambda,\delta,\mu ) = \\ &\mathbb{E}_i[ (1+\delta)R_1^{\text{NOMA}}(i) + (1+\mu)R_2^{\text{NOMA}}(i) \\& - \lambda(p_1(i)+p_2(i))] +\lambda\overline{P}-\delta\overline{R}-\mu\overline{R},
\end{split}
\end{equation}
where $ \lambda $ is the non-negative Lagrange multipliers for the constraint \eqref{objective:c1}, $\delta$ and $\mu$ are associated with constraints \eqref{objective:c4} for user 1 and user 2, respectively. Accordingly, the Lagrange dual function can be expressed as
\begin{equation}\label{dg1}
\begin{split}
&\mathcal{D}_{dt}^{\text{NOMA}}(\lambda,\delta,\mu) \\ &= \mathop{\text{max}}\limits_{ \{p_1(i),p_2(i)\}} { \mathcal{L}_{dt}^{\text{NOMA}}( \{ p_1(i) \}, \{ p_2(i) \}, \lambda,\delta,\mu )}.\\
\end{split}
\end{equation}
The dual of problem (P1.2) is then formulated as 
\begin{equation}\label{d1}
\begin{split}
\textbf{(D1)}\quad &\mathop{\text{min}}\limits_{ \lambda,\delta,\mu } {\quad \mathcal{D}_{dt}^{\text{NOMA}}},\\
&s.t. \quad \lambda \geq 0,\delta \geq 0,\mu \geq 0.
\end{split}
\end{equation}
The optimal solution for dual problem (D1) is equivalent to that for problem (P1.2), when the strong duality holds under time-sharing condition. In the following, we first obtain $\mathcal{D}_{dt}^{\text{NOMA}}(\lambda,\delta,\mu)$ from \eqref{dg1}, then decouple problem (D1) into multiple subproblems, and each subproblem denotes one fading state in the entire process. Consider one praticular fading state with given $(\lambda,\delta,\mu)$, the index $i$ can be ignored and the subproblem is given by 
\begin{equation}\label{ds1}
\begin{split}
\textbf{(D1.1)}\quad &\mathop{\text{max}}\limits_{ p_1 \geq 0, p_2 \geq 0 } {\quad \mathcal{\overline{L}}_{dt}^{\text{NOMA}}( p_1, p_2) } ,\\
&s.t. \quad p_1 + p_2 \leq \hat{P},
\end{split}
\end{equation}
where $\mathcal{\overline{L}}_{dt}^{\text{NOMA}}( p_1, p_2) = (1+\delta)R_1^{\text{NOMA}}+ (1+\mu)R_2^{\text{NOMA}}-
\lambda(p_1+p_2)$. Since all fading states are independent, all the subproblems can be solved parallelly. Therefore, we focus on solving problem (D1.1) in the sequel.  

\begin{proposition}
\emph{When $g_1 > g_2$, the optimal solution for problem (D1.1) is given by}
\begin{equation}\label{so1}
\begin{split}
\{p_1^*,p_2^*\}=arg max \{ &{\overline{L}}_{dt}^{\text{NOMA}}(0, 0), {\overline{L}}_{dt}^{\text{NOMA}}(0, \hat{P}),  \\ &{\overline{L}}_{dt}^{\text{NOMA}}(\hat{P}, 0),{\overline{L}}_{dt}^{\text{NOMA}}(p_{i,1}, p_{i,2}) \},
\end{split}
\end{equation}
\emph{where $(p_{i,1}, p_{i,2}), i \in \{1,2,3,4\}$ are the corresponding solution pair given as}
\begin{equation}\label{out,eq9}
\begin{cases}
p_{1,1}=0, \quad\qquad\qquad\qquad p_{1,2}=\left [ \frac{1+\mu}{\lambda ln2} - \frac{1}{g2} \right ]^{\hat{P}}_0,\\
p_{2,1}=\left [ \frac{1+\delta}{\lambda ln2} - \frac{1}{g1} \right ]^{\hat{P}}_0, \qquad p_{2,2}=0,\\
p_{3,1}=\left [ \frac{  (1+\mu)/g1 -(1+\delta)/g2 }{\delta-\mu} \right ]^{\hat{P}}_0,  \\ p_{3,2}=\left [ \hat{P}- \frac{  (1+\mu)/g1 -(1+\delta)/g2 }{\delta-\mu} \right ]^{\hat{P}}_0,\\
p_{4,1}= \frac{  (1+\mu)/g1 -(1+\delta)/g2 }{\delta-\mu},  \\ p_{4,2}= \frac{1+\mu}{\lambda ln2} - \frac{1}{g2}- \frac{  (1+\mu)/g1 -(1+\delta)/g2 }{\delta-\mu}.
\end{cases}
\end{equation}
\begin{proof}
Since $\overline{L}_{dt}^{\text{NOMA}}( p_1, p_2)$ is a continuous function over $\psi = \{(p_1 , p_2 )|p_1 \geq 0, p_2 \geq 0, p_1 + p_2  \leq \hat{P}\}$, its maximum proves to be either at the stationary point or on the boundary of $\psi$ when the stationary point is infeasible. The stationary point $(p_{4,1} , p_{4,2} )$ is given by
\begin{equation}\label{sd1}
(p_{4,1} , p_{4,2} )= arg\{ \nabla_{(p_1,p_2)}\overline{L}_{dt}^{\text{NOMA}}( p_1, p_2)=0 \}.
\end{equation}
If $(p_{4,1} , p_{4,2} ) \in \psi$, the maximum is obtained by ${\overline{L}}_{dt}^{\text{NOMA}}(p_{4,1}, p_{4,2})$, otherwise the maximum can be obtained by the boundary lies in $p_1=0, p_2=0$ or $p_1+p_2=\hat{P}$. Each boundary is denoted by $(p_{i,1}, p_{i,2}), i \in \{1,2,3\}$, respectively.
\end{proof}
\end{proposition}
We can find $\mathcal{D}_{dt}^{\text{NOMA}}$ efficiently by solving all sub-problems (D1.1) in parallel with given $(\lambda,\delta,\mu)$. The subgradient-based methods, such as the deep-cut ellipsoid method, can be applied to solve the dual problem \cite{boyd2004convex}. The problem (D1) sub-gradient for updating $(\lambda,\delta,\mu)$ is given by $(\overline{P}-\mathbb{E}_i[p_1^*(i)+p_2^*(i)]$, $\mathbb{E}_i[R_1^{*\text{NOMA}}(i)]-\overline{R}$, $\mathbb{E}_i[R_2^{*\text{NOMA}}(i)]-\overline{R})^T$, where $(p_1^*(i),p_2^*(i))$ is the optimal solution to (D1), and $R_k^{*\text{NOMA}}(i)$ is obtained by substituting the solution into \eqref{rate,eq3}.

The overall procedure for sovling the problem (P1) is summarized in Algorithm 3.

\begin{algorithm}
\caption{Iterative algorithm for problem (P1)}
\begin{algorithmic}[1]
\STATE \textbf{Initialize:} Feasible solutions $\{ p_k \}$ and phase shifter $\{ u(l) \}$;
\STATE Iteration count $l = 0$;
\REPEAT 
\STATE Solve problem (P1.2) with given $u$ and obtained the optimal $\{ p_k^{(l+1)}\}$.
\STATE Solve problem (P1.1.1) by algorithm 1 and obtained phase shifter $\{ u^{(l+1)}\}$. 
\STATE Obtain the discrete feasible solution $u^*$ via \eqref{rate,scaf}.
\STATE \textbf{if} objective value increases, \textbf{then}
\STATE $\{ u^{(l+1)}\} = u^*$;
\STATE \textbf{else} 
\STATE $u^{(l+1)} = u^{(l)}$;
\STATE \textbf{end}
\STATE $l=l+1$.
\UNTIL{The objective value convergence with the threshold $\epsilon > 0$.}
\end{algorithmic}
\end{algorithm}

\subsubsection{Optimal Solution to OMA Schemes} The power allocation problem for the OMA schemes is described as follows:
\begin{maxi!}|s|
{p_1(i),p_2(i)}{\mathbb{E}_v[R_1^{\text{OMA}}(i)+R_{2}^{\text{OMA}}(i)]}
{}{\textbf{(P2.3)}}
\addConstraint{\eqref{objective:c1}, \eqref{objective:c2}, \eqref{objective:c3}, \eqref{objective:c4}.}{}
\end{maxi!}

The non-convexity still lies in problem (P2.3), since $\alpha_k(i)log_2(1+p_k(i)g_k(i))$ is jointly concave with respect to $\alpha_k(i)$ and $p_k(i)$. Then, for the strong duality, the Lagrangian dual method is applied for solving problem (P2.3) and the Lagrangian function is given by 
\begin{equation}\label{L2}
\begin{split}
\mathcal{L}_{dt}^{\text{OMA}}( & \{ p_1(i) \}, \{ p_2(i) \},\{ \alpha_1(i) \}, \lambda,\delta,\mu )= \\ &\mathbb{E}_i[ (1+\delta)R_1^{\text{OMA}}(i) + (1+\mu)R_2^{\text{OMA}}(i) \\ & -\lambda(p_1(i)+p_2(i))] +\lambda\overline{P}-\delta\overline{R}-\mu\overline{R},
\end{split}
\end{equation}
where $ \lambda $, $\delta$ and $\mu$ are the non-negative Lagrange multipliers similar with those in NOMA scheme, and $\alpha_2 = 1 - \alpha_1$. Similarly, we can decouple $\mathcal{L}_{dt}^{\text{OMA}}( \{ p_1(i) \}, \{ p_2(i) \},\{ \alpha_1(i) \}, \lambda,\delta,\mu )$ into parallel sub-Lagrangian for the same structure, with $\mathcal{\overline{L}}_{dt}^{\text{OMA}}(p_1, p_2, \alpha_1) = (1+\delta)R_1^{\text{OMA}}+ (1+\mu)R_2^{\text{OMA}}-\lambda(p_1+p_2)$, where the fading state index $i$ has been ignored as sub-problems are considered in each fading state. Then, one particular sub-problem corresponding a single fading state is formulated as
\begin{equation}\label{ds2}
\begin{split}
\textbf{(D2)}\quad &\mathop{\text{max}}\limits_{ p_1 \geq 0, p_2 \geq 0, \alpha_1 } {\quad \mathcal{\overline{L}}_{dt}^{\text{OMA}}( p_1, p_2,\alpha_1) } ,\\
&s.t. \quad\quad p_1 + p_2 \leq \hat{P},\\
&\quad\quad\quad\quad  0\leq \alpha_1 \leq 1.
\end{split}
\end{equation}
\begin{lemma}
\emph{If the maximun of $\mathcal{\overline{L}}_{dt}^{\text{OMA}}( p_1, p_2,\alpha_1)$ is achieved by the jointly stationary point, the following conditions should be satisfied:}
\begin{subequations}  \label{eq:1}
\begin{align}   
&h(\lambda,\theta,\mu)=0,            \label{eq:1A} \\
&c_1 \leq 0,            \label{eq:1B} \\
&c_2 \leq 0,           \label{eq:1C} \\
&\begin{cases}
\frac{ \hat{P}-c_2 }{c_1-c_2}\leq 0, \quad if \quad c_1 \leq c_2,   \\
\frac{ \hat{P}-c_2 }{c_1-c_2}\geq 1, \quad otherwise,  \label{eq:1D}
\end{cases}
\end{align}
\end{subequations}
\emph{where $c_1=\frac{1+\delta}{\lambda ln2}-\frac{1}{g_1}$, $c_2=\frac{1+\mu}{\lambda ln2}-\frac{1}{g_2}$ and $h(\lambda,\theta,\mu)$ is given by}
\begin{equation}\label{h}
\begin{split}
h(\lambda,\theta,\mu)= & (1+\delta)log_2(\frac{1+\delta}{\lambda ln2}g_1) \\
                       & - (1+\mu)log_2(\frac{1+\mu}{\lambda ln2}g_2)-\lambda c_1+ \lambda c_2.
\end{split}
\end{equation}

\emph{The stationary point is thus given by} 
\begin{subequations}  \label{eq:2}
\begin{align}   
&p_1^*=c_1 \alpha_1^*, \quad p_2^*=c_2 \alpha_2^*           \label{eq:2A} \\
&\alpha_1^* =
\begin{cases}
\forall \in [ 0 , min \{ \frac{ \hat{P}-c_2 }{c_1-c_2} , 1 \} ], if \quad c_1 > c_2  \\
\forall \in [ ( \frac{ \hat{P}-c_2 }{c_1-c_2})^+ , 1 \} ], \quad otherwise.  \label{eq:2D}
\end{cases}
\end{align}
\end{subequations}
\begin{proof}
The jointly stationary point is achieved by solving $\nabla_{(p_1,p_2,\alpha_1)}\overline{L}_{dt}^{\text{OMA}}( p_1, p_2,\alpha_1)=0$, and \eqref{eq:1A} is obtained by plugging $p_1=c_1\alpha_1$ and $p_2=c_2\alpha_2$ into the partial derivative $\overline{L}_{dt}^{\text{OMA}}( p_1, p_2,\alpha_1)$ with respect to $\alpha_1$. Then, considering $p_1\leq 0$, $p_2 \leq 0$ and $p_1+p_2 \geq \hat{P}$, we have \eqref{eq:1B}, \eqref{eq:1C} and \eqref{eq:1D} denotes the feasible range for $\alpha_1$, respectively.
\end{proof}
\end{lemma}
If $\overline{L}_{dt}^{\text{OMA}}( p_1, p_2,\alpha_1)$ achieves the maximum on the boundary points where $p_1+p_2=\hat{P}$ , the optimum $( p_1, p_2,\alpha_1)$ should be
\begin{equation}\label{h1}
\begin{cases}
p_1^*=0, p_2^*=\hat{P}, \alpha_1^*=0, \quad if \quad \frac{1+\mu}{1+\delta}>\frac{log_2(1+\hat{P}g_1)}{log_2(1+\hat{P}g_2)}, \\
p_1^*=\hat{P}, p_2^*=0, \alpha_1^*=1, \quad otherwise.
\end{cases}
\end{equation}

The optimal soultion to problem (D2) is thus given by
\begin{equation}\label{h2}
\begin{split}
(p_1^*, p_2^*, \alpha_1^*) = &argmax\{ \mathcal{\overline{L}}_{dt}^{\text{OMA}}(0,0,0),\\ &\mathcal{\overline{L}}_{dt}^{\text{OMA}}(0,\hat{P},0), \mathcal{\overline{L}}_{dt}^{\text{OMA}}(\hat{P},0,1),\\ &\mathcal{\overline{L}}_{dt}^{\text{OMA}}(0,c_2,0), \mathcal{\overline{L}}_{dt}^{\text{OMA}}(c_1,0,1)
\} ,
\end{split}
\end{equation}
Problem (D2) is solved by solving the subproblems with given $(\lambda,\delta,\mu)$. The sub-gradient method is also applied with the corresponding sub-gradient $(\overline{P}-\mathbb{E}_v[p_1^*(i)+p_2^*(i)], \mathbb{E}_v[R_1^{*OMA}(i)]-\overline{R}, \mathbb{E}_v[R_2^{*OMA}(i)]-\overline{R})^T $ for updating $(\lambda,\delta,\mu)$. Then, problem (P2.3) is iteratively solved and the overall algorithm for the IRS-aided OMA scheme is summarized in Algorithm 4.
\begin{algorithm}
\caption{Iterative algorithm for problem (P2)}
\begin{algorithmic}[1]
\STATE \textbf{Initialize:} Feasible solutions $\{ p_k \}$ and phase shifter $\{ u(l) \}$;
\STATE Iteration count $l = 0$;
\REPEAT 
\STATE Solve problem (P2.2) with given $u$ and obtained the optimal $\{ p_k^{(l+1)}\}$.
\STATE Solve problem (P2.1) or problem (P2.2) obtained phase shifter $\{ u^{(l+1)}\}$. 
\STATE Obtain the discrete feasible solution $u^*$ via \eqref{rate,scaf}.
\STATE \textbf{if} objective value increases, \textbf{then}
\STATE $\{ u^{(l+1)}\} = u^*$;
\STATE \textbf{else} 
\STATE $u^{(l+1)} = u^{(l)}$;
\STATE \textbf{end}
\STATE $l=l+1$.
\UNTIL{The objective value convergence with the threshold $\epsilon > 0$.}
\end{algorithmic}
\end{algorithm}
\subsection{Complexity and Convergence} 

The AO method is applied for problems (P1) and (P2) and detailed in Algorithms 3 and 4. The proposed phase shifters adjusment algorithms contain interior-point method, and the complexity is given by $\mathcal{O}(I^s(3k^2+N^2)^{3.5})$, where $I^s$ defines the iteration number of the SROCR method in Algorithms 1 and 2. The convergence of Algorithms 1 and 2 is similar and we take Algorithm 1 for example. With given power, time and frequency allocations for problem (P1.1), we have
\begin{equation}\label{c2}
\begin{split}
x(\{ p_k^l \}, u^l) &\overset{(a)}= x^{lb}(\{ p_k^l \}, u^l)\\ 
& \overset{(b)}\leq x^{lb}(\{ p_k^{l+1} \}, u^{l+1})\\
& \overset{(c)}\leq x(\{ p_k^{l+1} \}, u^{l+1}) ,
\end{split}
\end{equation}
where $x^{lb}$ denotes the problem (P1.1.2)’s value in l-th iteration. With given local points in problem (P1.1.2), the first-order Taylor expansions are tight and (a) holds. Due to "time-sharing" condition is satisfied, problem (P1.2) is solved with optimal solution and (b) holds. Since relaxation replacement for rank-one constrain, (c) holds. As a result, the obtained problem (P1.1)’s value remains non-decreasing in each iteration. Then, we have
\begin{equation}\label{c3}
x(\{ p_k^l \}, u^l) \leq x(\{ p_k^{l+1} \}, u^{l+1}).
\end{equation}


\begin{remark}\label{re2}
The maximum is bounded by a finite value. The proposed algorithms always converge to a locally optimal solution for both problems (P1) and (P2) as shown in Equation \eqref{c3}.
\end{remark}

\begin{figure}[H]
\centering
\includegraphics[width= 2.5in]{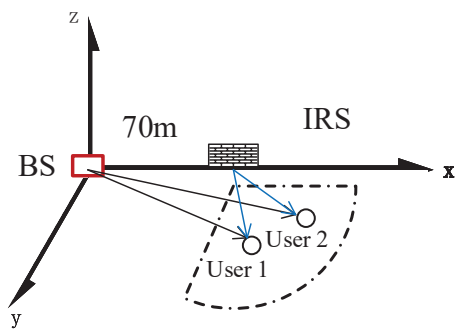}
\caption{The simulated IRS-assisted network.}
\label{dis}
\end{figure}

\section{Simulation Results}

In this section, numerical results are persented for verifying our proposed algorithms. The downlink IRS-assisted system over fading channels is illustrated in Fig. 2. We set the IRS phase quantization level as $L=3$. The BS and the IRS's locations are at (0, 0, 0) and (70, 0, 0), respectively. The IRS is located between the BS and the $K$ served users, which are randomly distributed in a $1/4$ circle region centered at (80, 10, 0) with a radius of 40 m. The distance from BS to IRS $d_i$ and that from IRS to the users $d_u$ follows the relation $d_i + d_u \leq 120m$, where the IRS is capable of acting as an anomalous mirror. The Rayleigh fading channel model is modeled for the direct link and the rician fading model is modeled for the assistant link. Let $v_{BU}$ and $v_{UI}$ denote the Rician factors of the BS-IRS and IRS-user links, where $v_{BI}=v_{IU}=3dB$. The path loss exponents for the BS-user, BS-IRS, and IRS-user links are set to be $\varphi_{BU}=3.5$, $\varphi_{BI}=2.2$ and $\varphi_{BU}=2.8$, respectively. The convergence threshold is $\xi=10^{-2}$ and the noise power is set as $\sigma^2=-90dBm$. The number of fading states is set as $F = 10^7$ as an approximation to infinity.

\begin{figure}[H]
\setlength{\abovecaptionskip}{0.cm}
\setlength{\belowcaptionskip}{0.cm}
\centering
\includegraphics[width= 3.6in, height=2.2in]{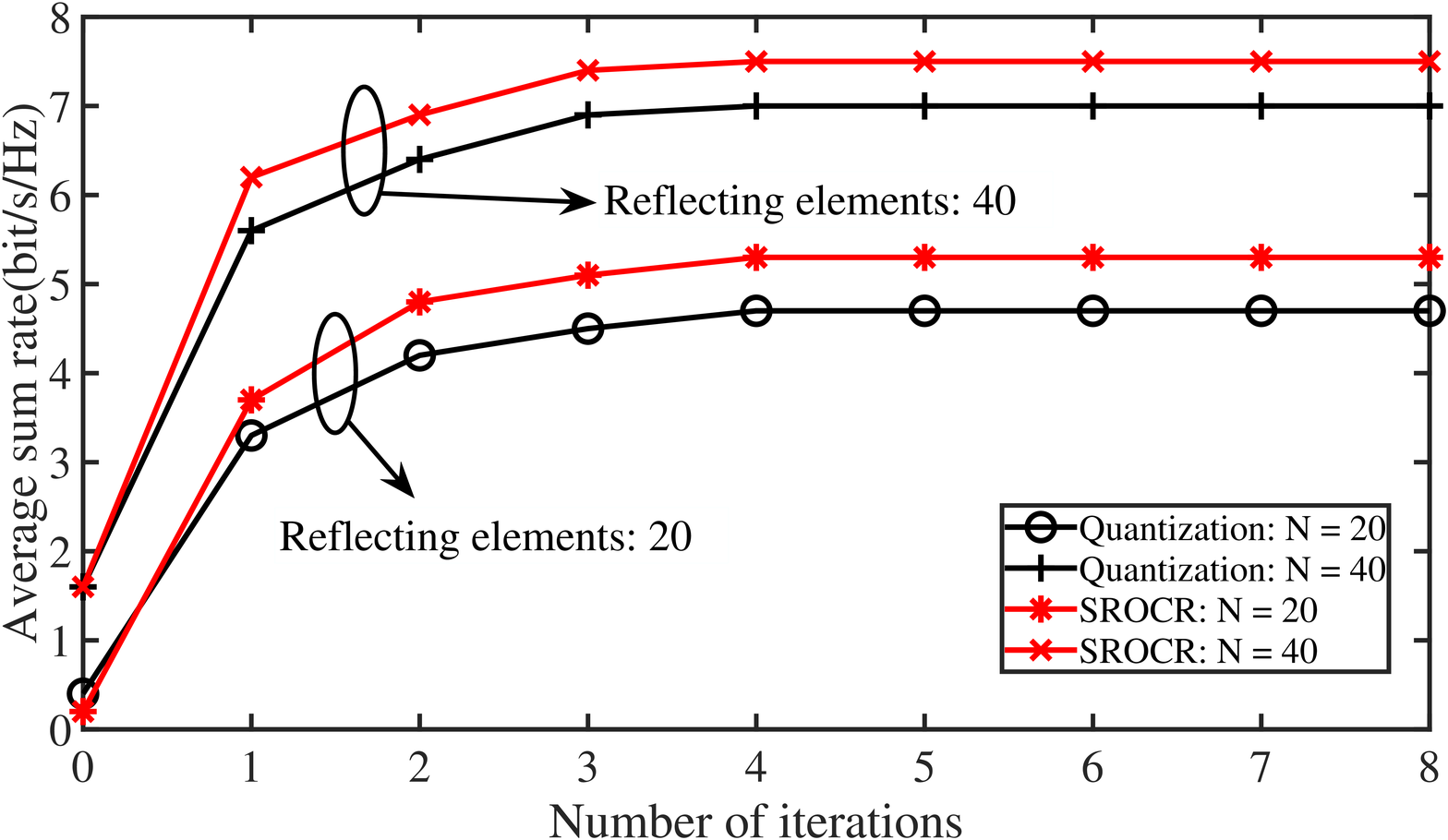}
\caption{The average sum rate versus number of iterations.}
\label{iteration}
\end{figure}
\textbf{The convergence of the SROCR with/without quantization versus iteration numbers:} Fig. \ref{iteration} illustrates the achieved sum rate versus the number of iterations required by proposed algorithms. We first randomly generate continuous phase shifts and the quantize them to its nearest discrete value set as the initial solution. The SROCR is applied to obtain the continuous solution, where the relaxation parameter $\kappa \in [0,1]$ determines the relaxation levels from dropping the rank-one constrain to applying the original constrain. Then we obtain the discrete phase shifts by quantizing. Fig \ref{iteration} shows that with a small number of iterations, proposed algorithms can converge, which verifies the insights gleaned from \textbf{Remark \ref{re2}}. Thus we can build a feasible rank-one solution efficiently.

\begin{figure}[H]
\setlength{\abovecaptionskip}{0.cm}
\setlength{\belowcaptionskip}{0.cm}
\centering
\includegraphics[width= 3.6in, height=2.2in]{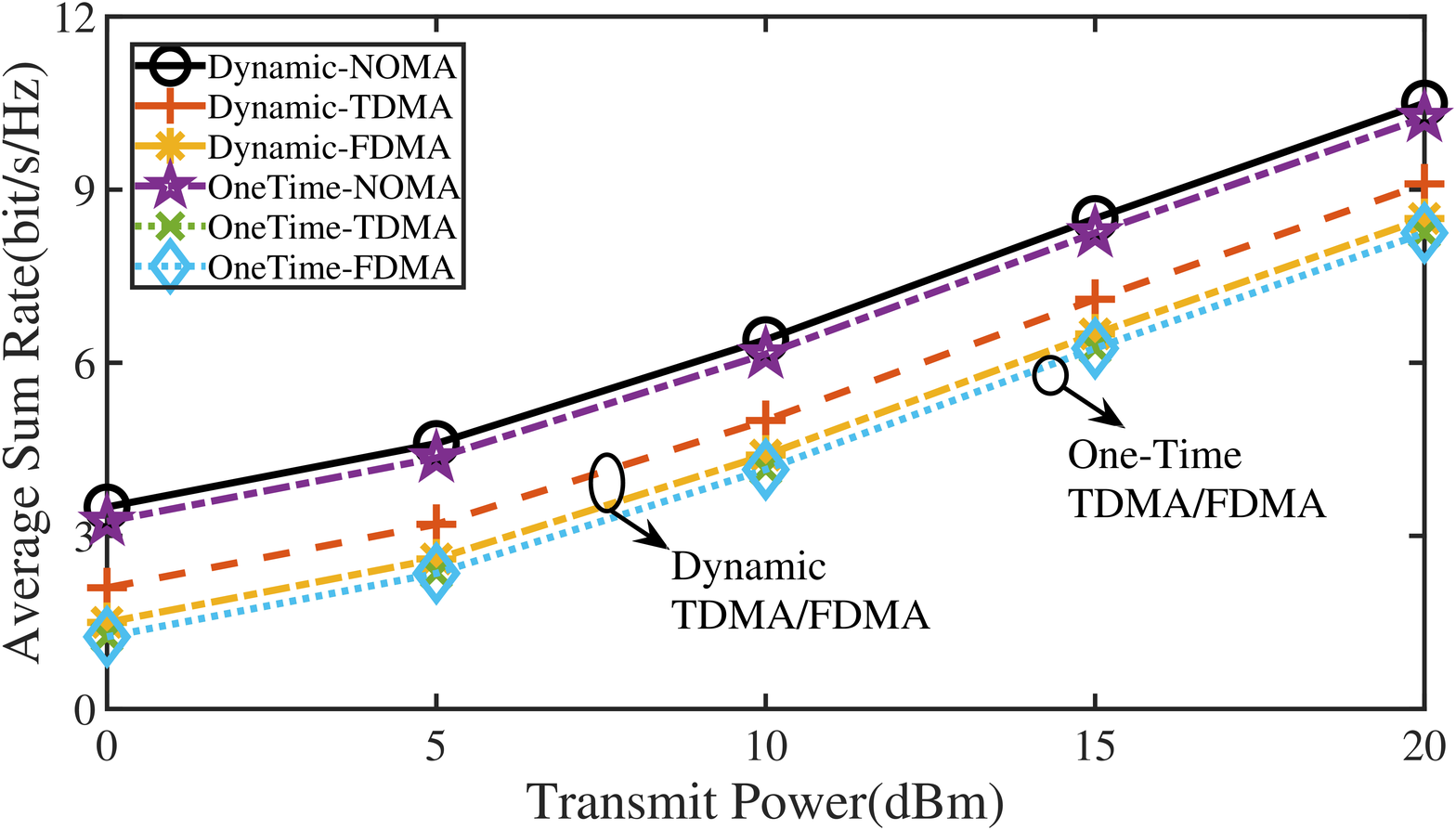}
\caption{The average sum rate versus transmit power under various phase adjustment.}
\label{Dynamic}
\end{figure}
\textbf{Average sum rate versus transmit power under various phase adjustment:} As shown in Fig. \ref{Dynamic}, the performence in the dynamic phase adjustment scheme outperforms the one-time phase adjustment scheme, as the reflceting elements are optimised in every fading state. The gain of adjusting the phase shift in every fading state is not pronounced while the complexity and the power for adjusting the reflecting elements increases significantly. The IRS-assisted NOMA scheme outperforms OMA scheme under both dynamic phase adjustment and one-time phase adjustment. It should be noted that the user deployment is asymmetric, which is to ensure the successful implementation of SIC in the NOMA scheme. In addition, for the hardware limitations of the IRS elements, IRS has the feature of time-selective, while frequency-selective can not be achieved by the phase shifters. The TDMA scheme outperforms the FDMA scheme under dynamic phase adjustment for the proposed system, which validates our \textbf{Remark \ref{re1}}.

\begin{figure}[H]
\setlength{\abovecaptionskip}{0.cm}
\setlength{\belowcaptionskip}{0.cm}
\centering
\includegraphics[width= 3.6in, height=2.2in]{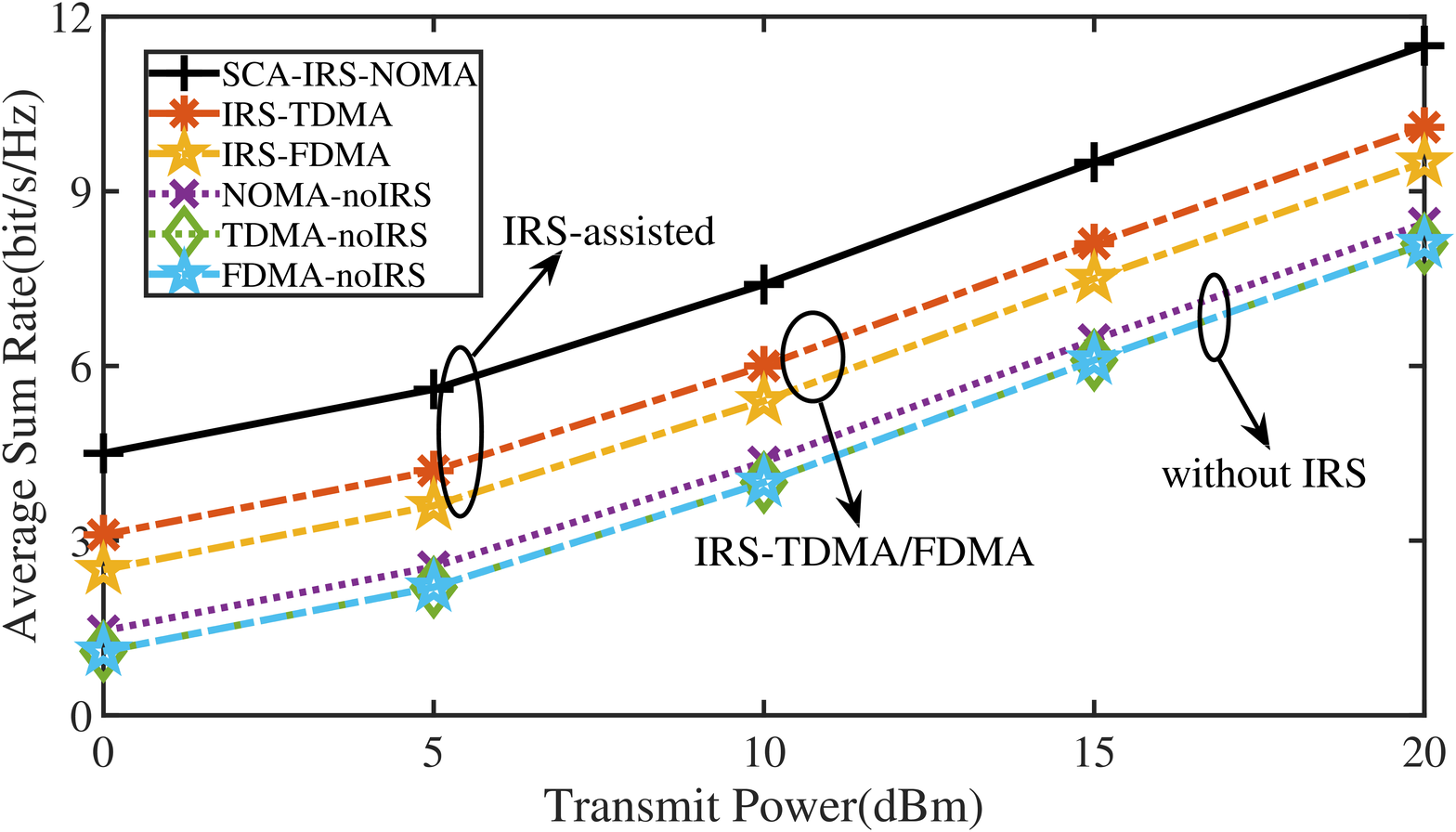}
\caption{The average sum rate versus transmit power under dynamic phase adjustment.}
\label{TFDMA}
\end{figure}
\textbf{Average sum rate versus transmit power under dynamic phase adjustment:} As shown in Fig. \ref{TFDMA}, the system average sum rate versus the power budget under different schemes. The number of IRS elements is set to $N=30$. Under various multiple access schemes, the average sum rate increases with the higher transmit power. Compared to the traditional NOMA scheme, the performence gain comes from the enhanced combined-channel introduced by the IRS. Moreover, larger channel condition differences can be achieved for the IRS has the ability to adjust the propagation environment. The IRS can also enhance the coverage for the additional links established, which makes the NOMA pairs scheduled with poor direct links or close channel gains to achieve better performance. It is even possible to adjust the decoding order in NOMA scheme with the IRS reconfiguration, in some scenarios that have primary user to serve. The ability of steering signals towards certain directions birngs performance enhancement when the target rates of user are different. Besides, NOMA scheme can achieve higher average throughput in comparison with the OMA scheme. As NOMA allows mutiple users to share the same resource block, the higher spectral efficiency can be obtained. 

\begin{figure}[H]
\setlength{\abovecaptionskip}{0.cm}
\setlength{\belowcaptionskip}{0.cm}
\centering
\includegraphics[width= 3.6in, height=2.2in]{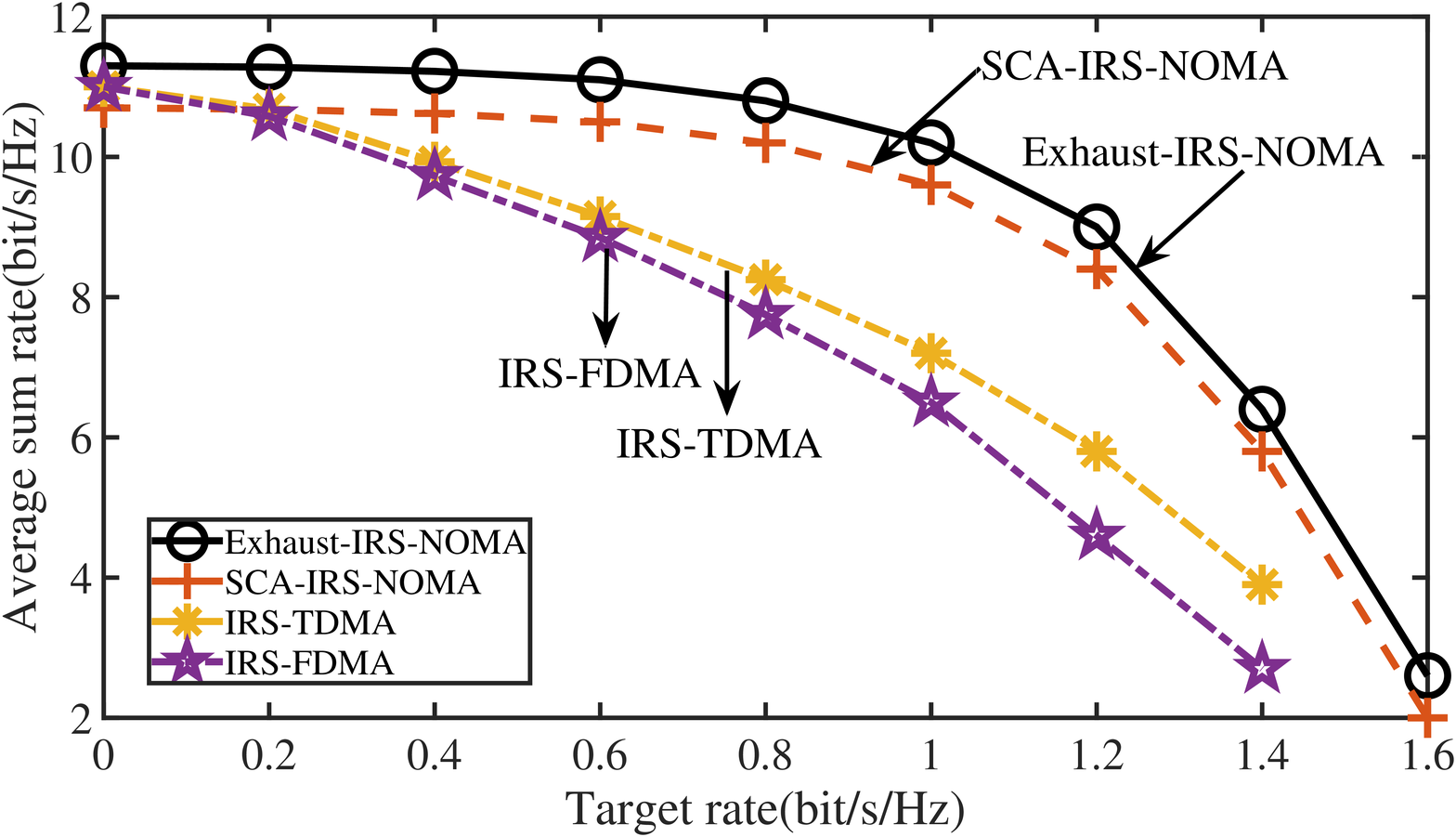}
\caption{The average sum rate versus users target rate under dynamic phase adjustment.}
\label{target}
\end{figure}
\textbf{Average sum rate versus users target rate under dynamic phase adjustment:} Fig. \ref{target} depicts the relationships between the average rate performance of the network and the users fairness, obtained form NOMA and OMA schemes assisted by IRS. {The target rate is set to ensure the system fairness, which guarantees the data rate for the weak user. It can be seen that the IRS-aided NOMA network outperforms the IRS-aided OMA network under various target rate settings. The gap reduces when the target rate is very small since most resources are allocated to the user with better combined channel gain. IRS-assisted NOMA scheme is seen to be more robust against the target rate increase due to its ability of using the single resource block to serve multiple users. We can also see that TDMA scheme outperforms FDMA scheme as the IRS can be adjusted to maximise each user's combined-channel in the TDMA scheme under dynamic phase adjustment.

\begin{figure}[H]
\setlength{\abovecaptionskip}{0.cm}
\setlength{\belowcaptionskip}{0.cm}
\centering
\includegraphics[width= 3.6in, height=2.2in]{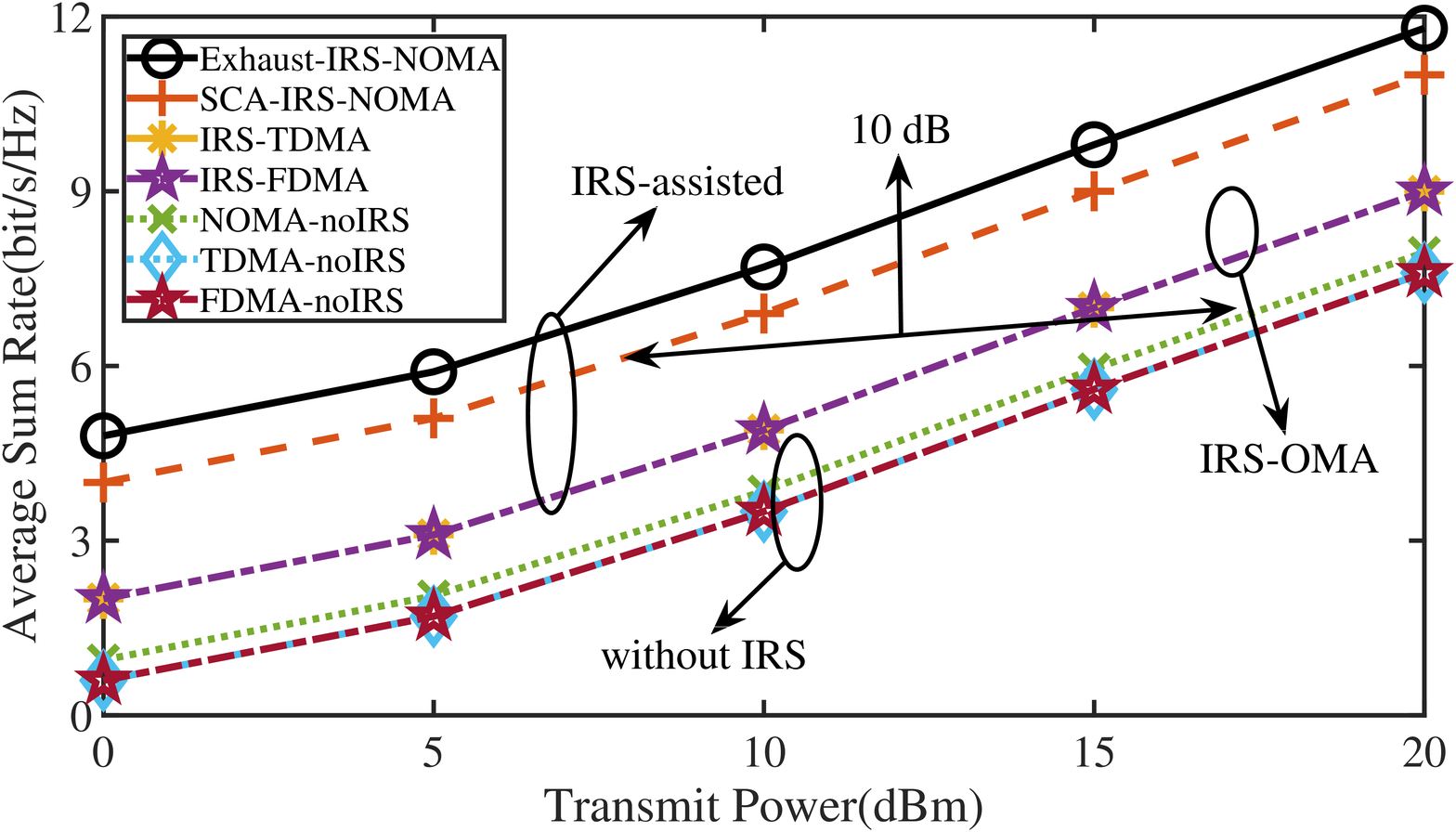}
\caption{The average sum rate versus power budget under one-time phase adjustment.}
\label{power}
\end{figure}
\textbf{Average sum rate versus power budget under one-time phase adjustment:} Fig.\ref{power} presents the proposed algorithms performances can obtain the close average sum rate to that obtained by the brutal search with lower complexity. The feasible solution in the discrete phase shifters adjustment that we achieve may experience performance losses for the quantization method. Appling IRS brings a 10 dB power consumption gain compared to the traditional NOMA scheme. The IRS-assisted NOMA scheme achieves higher average sum rate than traditional mutilple access schemes while ensuring the system fairness. Moreover, in one-time adjustment scheme, the TDMA scheme has the same performence as the FDMA scheme as the reflecting elements can only be adjusted once at the nodes of fading blocks. 


\begin{figure}[H]
\setlength{\abovecaptionskip}{0.cm}
\setlength{\belowcaptionskip}{0.cm}
\centering
\includegraphics[width= 3.6in, height=2.2in]{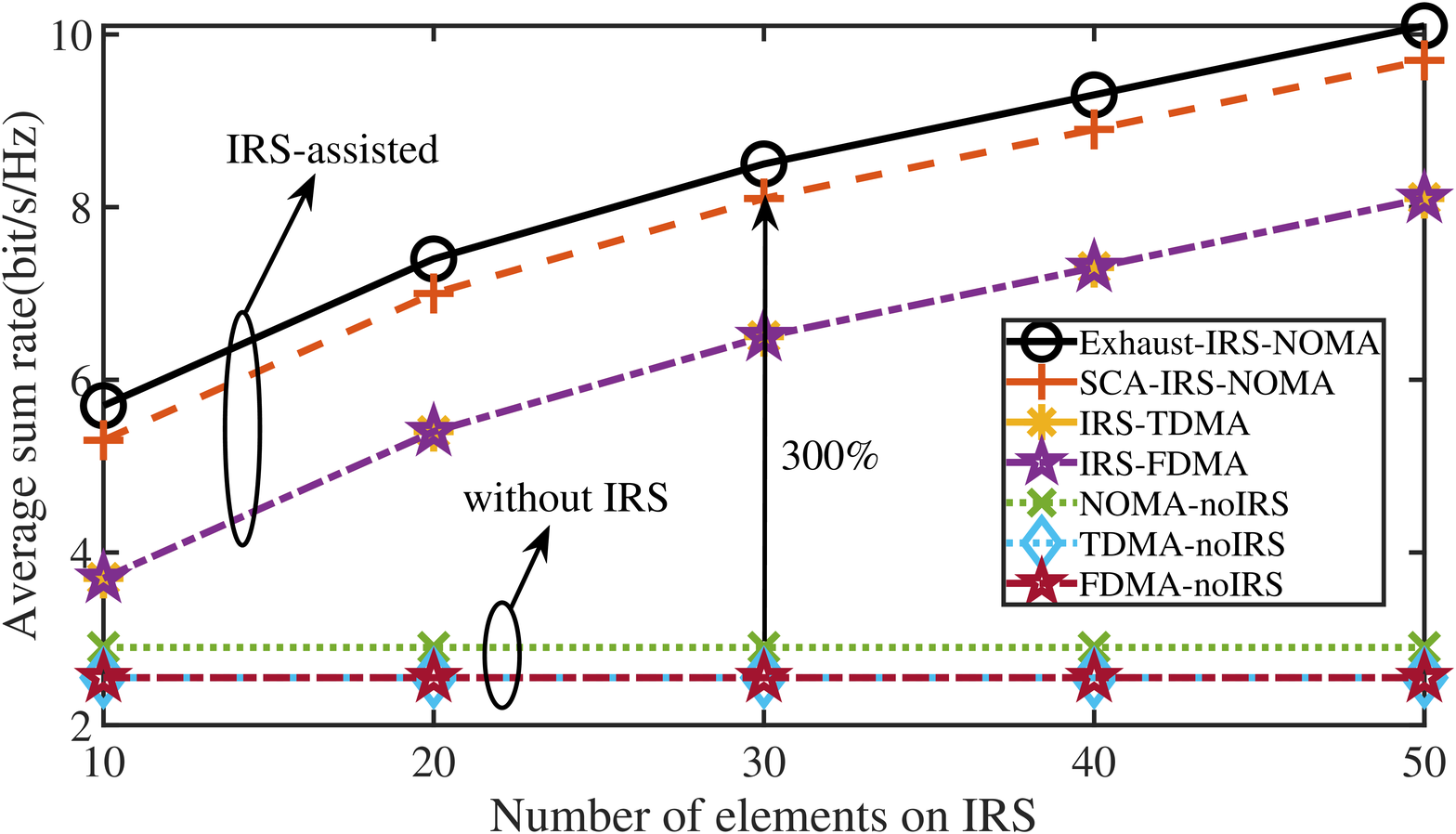}
\caption{The average sum rate versus number of IRS elements under one-time phase adjustment.}
\label{number}
\end{figure}
\textbf{Average sum rate versus the number of IRS elements under one-time phase adjustment:} In Fig. \ref{number}, the comparison between system average data rate of various mutilple access schemes versus the amount of reflecting elements N is presented. We can observe that IRS aided schemes outperforms the other schemes without IRS. Moreover, with the increase on phase elements $N$, the system performance is further enhanced. The larger number of refleting elements gives flexibility on phase shifter adjustment and leads to higher combined channel gains. The recieved power level can be significanly increased with more refleting elements, as more signals' power is reflected to served users.

\begin{figure}[H]
\setlength{\abovecaptionskip}{0.cm}
\setlength{\belowcaptionskip}{0.cm}
\centering
\includegraphics[width= 3.6in, height=2.2in]{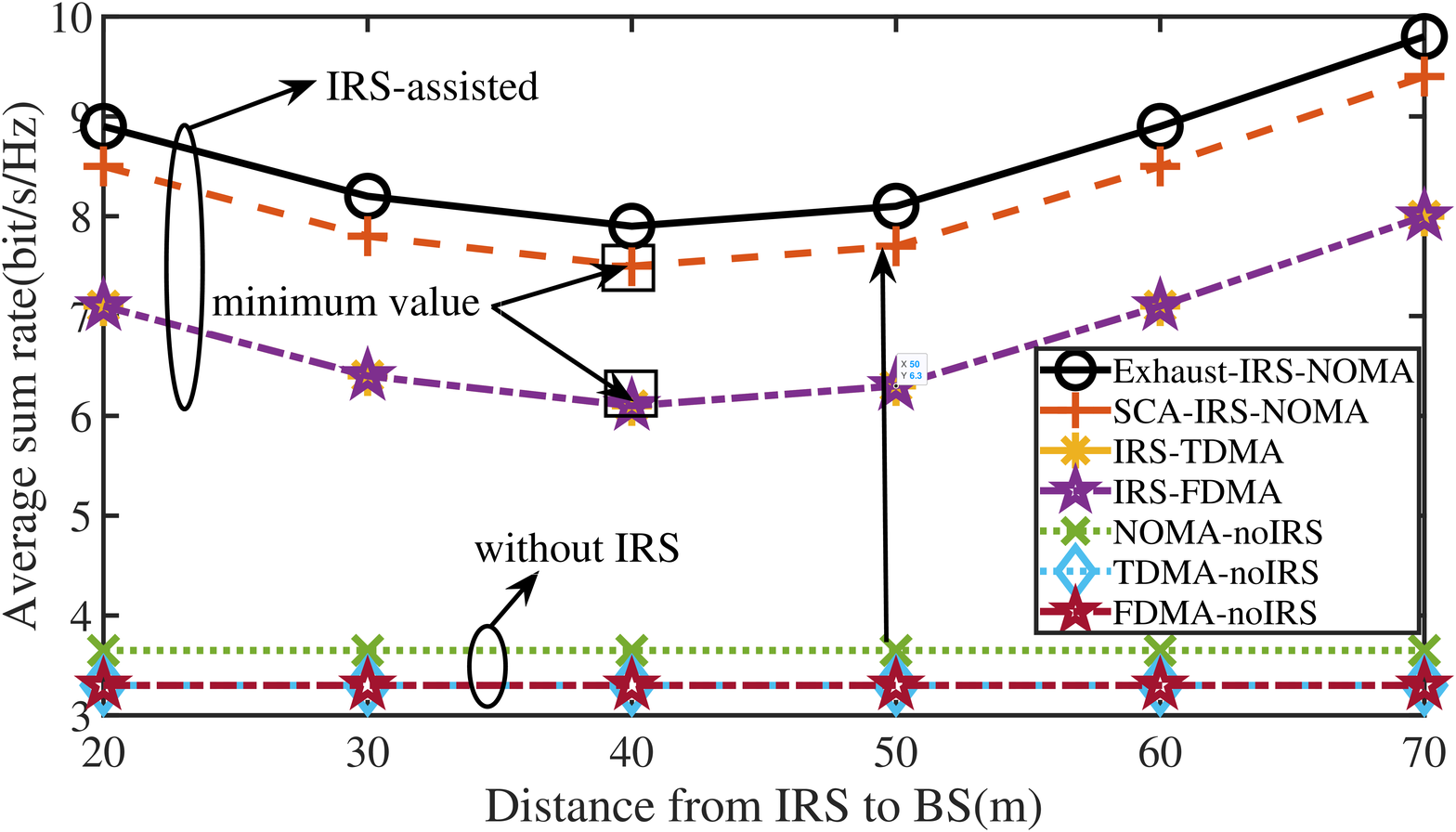}
\caption{The average sum rate versus distance from IRS to BS under one-time phase adjustment.}
\label{distance}
\end{figure}
\textbf{Average sum rate versus distance from IRS to BS under one-time phase adjustment:} As shown in Fig. \ref{distance}, the average rate performance of the IRS-assisted systems first decrease when the IRS is deployed close to the BS. Observe that the minimum data rate is reached when the IRS is deployed in the middle between the BS and users, where $d_i$ is similar to $d_u$, as such deployment will decrease the amplitude of the concatenated channel gains. The average sum rate will be enhanced by a better IRS-aided link achieved from the shorter distance between the IRS and the user. The large number of IRS elements and bits for phase shift quantization can also enhance the average sum rate and the required phase shift quantization bits decrease as the number of IRS elements grows. While more IRS elements and bits for phase shift quantization also bring the computational complexity and implement cost, and the limited performance degradation is observed between the infinite number of elements and dozens of elements\cite{zhang2020reconfigurable}. The orders in NOMA scheme are also critical when IRS can influence the channel gains. Thus, with proper IRS elements, bits for phase shift quantization, orders in NOMA scheme, and choice of the IRS location, the system performance could be enhanced.

\section{Conclusion}

The IRS has been applied in the downlink transmission with two users over fading channels for enhancing the average sum rate. Specifically, IRS assisted NOMA and OMA schemes were investigated. Additionally, we divided all fading states into several fading blocks and the phase shifters were adjusted at the nodes of the fading blocks. For phase shift design, we proposed an effective solution using SCA to obtain high quality solutions. In addition, a SROCR method was applied to deal with the rank-one constraint. Then, we obtain locally optimal continuous phase shifters. The discrete phase shifters were obtained by a quantization scheme. For power allocation in each fading state, we solved the non-convex optimization problems using the dual decomposition method leveraging “time-sharing” conditions. Then, the average sum rate was maximized via the Lagrangian dual decomposition. Simulation results have revealed that in the downlink system over fading channels, the IRS assisted NOMA scheme can enhance the system performance significantly. The multiple antennas and multiple users could lead to potential performance gain aided by IRS, hence, our future research is to investigate decoding order and user pairing in systems with multiple antennas.

\bibliographystyle{IEEEtran}
\bibliography{bib2018}

\end{document}